\documentclass[aps,prl,groupedaddress,showpacs,amsmath,twocolumn,amssymb,floatfix]{revtex4-1}
\usepackage{graphicx}
\usepackage{bm}
\usepackage{color}
\usepackage{epstopdf}

\newcommand{\he}{$^3$He}
\newcommand{\hefour}{$^4$He}
\newcommand{\she}{superfluid $^3$He}

\newcommand{\Bph}{{\it B} phase}
\newcommand{\Aph}{{\it A} phase}

\begin{document}

\title{The Effect of Magnetic Impurities on Superfluid \he\ in Aerogel}

\author{A.M. Zimmerman}
\email[]{andrewzimmerman2016@u.northwestern.edu}
\author{M.D. Nguyen}
\author{J.W. Scott}
\author{W.P. Halperin}
\email[]{w-halperin@northwestern.edu}
\affiliation{Northwestern University, Evanston, IL 60208, USA}

\date{\today}

\begin{abstract}
The critical field for \she\ in axially compressed, anisotropic silica aerogel is shown to be the result of an anisotropic distribution of magnetic impurities affecting the superfluid $A$ phase.  The critical field results from the fact that the $A$ phase  is suppressed relative to the $B$ phase which is immune to the effects of magnetic impurities.  In the absence of magnetic quasiparticle scattering in anisotropic aerogel, we find that the relative symmetry of $A$ and $B$ phase order parameters is the same as in isotropic aerogel, just as it is in pure \she.   These results are of potential  importance for understanding unconventional superconductivity.
\end{abstract}

\maketitle

Unlike conventional superconductors, unconventional superconductors and superfluids are affected by the presence of both magnetic and non-magnetic impurity scattering of quasiparticles. These effects can be investigated using superfluid $^3$He, a well-established unconventional superfluid.  Pure \he\ has two unique $p$-wave, spin triplet superfluid states at low magnetic fields: the isotropic \Bph\ and, above a pressure of 21 bar, the anisotropic \Aph\ \cite{Vol.90}. 
Although \he\ is inherently pure, highly porous  aerogel can be used to introduce impurity into the system \cite{Hal.18}. Aerogels consist of networks of small strands that act as impurities. The structure of these aerogel strands has  a large effect on the resulting superfluid phase diagram due to the presence or absence of global anisotropy. 
Anisotropy of the \he\ quasiparticle mean-free-path results from a preferred direction for the strands, playing a large role in the stability of phases with different order parameter symmetry~\cite{Pol.12a, Li.14a, Li.14b, Li.15, Zhe.16, Asa.15, Dmi.15, aut.16, Dmi.18}.  Here, we report  the significant effects of anisotropic magnetic impurities on phase stability.

Recent experiments have been conducted in highly anisotropic, nematically ordered alumina aerogels in which the aerogel strands are nearly parallel \cite{Asa.15,Dmi.15,DMi.15b,Dmi.18}. 
In the presence of these highly ordered impurities, new  physical phenomena have been reported, including a new superfluid phase, the Polar phase \cite{Dmi.15}, and half-quantum vortices \cite{aut.16}. The pressure-temperature superfluid phase diagram in this system appears to be greatly affected by magnetic impurities~\cite{Dmi.18}, raising the question of how general this phenomenon is. To answer this question, we investigate the role of magnetic impurities on superfluid $^3$He in anisotropic silica aerogels.

\begin{figure}

\centerline{\includegraphics[width=0.45\textwidth]{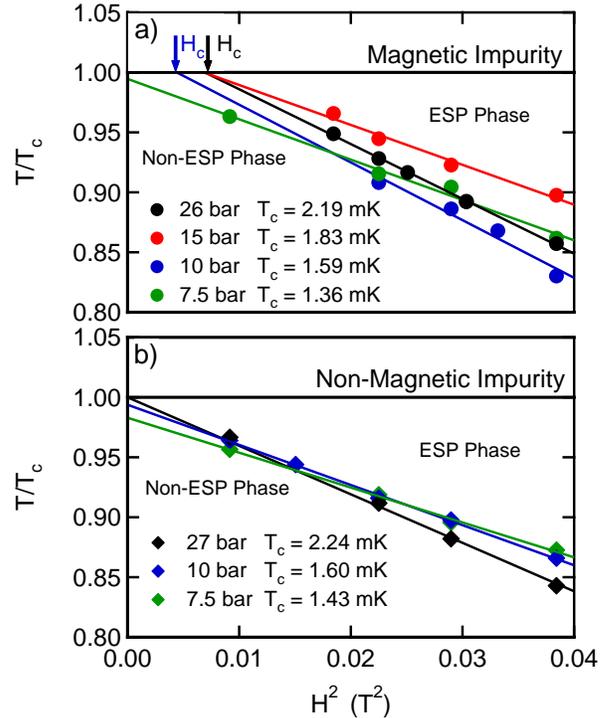}}

		\caption{\label{fig:H_sq}(Color online). The temperature-field superfluid phase diagram showing the dependence of $T_{AB}/T_c$ on magnetic field at a variety of pressures. $T_{AB}/T_c$ depends quadratically on magnetic field in all cases. a) With magnetic impurities a critical field, $H_c$, is present at $P=27$\,bar ($H_c=88.6$\,mT) 15 bar ($H_c=82.6$\,mT), and 10 bar ($H_c=66.4$\,mT) in the presence of magnet scattering, but is absent at lower pressure. b) With non-magnetic impurity $H_c=0$.}

\end{figure} 

Aerogels used in experiments on \she\ are not intrinsically magnetic; however, a few layers of paramagnetic solid \he\ adsorbed on the surface creates a channel for magnetic quasiparticle scattering \cite{Sau.05, Col.09, Sch.87}. This paramagnetic solid can be removed by replacing the magnetic \he\ on the surface with non-magnetic \hefour, allowing the switch from magnetic to non-magnetic impurity. Almost all of our earlier work has been with magnetic aerogel impurities, with the  exception of Sprague {\it et al.}~\cite{Spr.96}. We note that the addition of \hefour\ also modifies the specularity of quasiparticle scattering, although this effect should be negligible at high pressures \cite{Fre.90,Tho.91,Kim.93,Mur.12}.
Dmitriev {\it et al.}~\cite{Dmi.18} show that the newly observed Polar phase  is only present with non-magnetic aerogel impurities.  Additionally, the transition temperature, $T_c$,  from the normal state to the superfluid was noticeably suppressed~\cite{Dmi.18}. The authors suggested that the large influence of magnetic scattering might be due to the large anisotropy of the impurity itself. This effect was not observed in early experiments with isotropic aerogels~\cite{Dmi.03, Spr.96}.

Most theoretical work has not addressed the effects of magnetic impurities \cite{Thu.98,Sau.13,Aoy.06,Aoy.05}, or focused on  magnetic impurities in the absence of anisotropy  \cite{Sau.03, Bar.00,Ike.09}.
New calculations, motivated by Ref.\cite{Dmi.18}, indicate that magnetic impurity might reduce the effects of anisotropy \cite{Min.18}. However other recent calculations find  only small changes in the phase diagram due to magnetic scattering \cite{Fom.18}. Clearly, more experimental work is needed.

We  measured the pressure-temperature-field phase diagram of \she\ with magnetic and non-magnetic impurities using an aerogel sample with less anisotropy than alumina aerogels~\cite{Dmi.18}, and found that the phase diagram is again significantly modified by magnetic impurity. In particular, the superfluid \Aph\ appears to be suppressed, much as has been reported in the work on the Polar phase~\cite{Dmi.18}. It is interesting to note that both the Polar and \Aph s are anisotropic equal spin pairing (ESP) states with the same magnetic susceptibility as the normal state; however, only the \Aph\  breaks time-reversal symmetry. Unlike in measurements in nematic aerogel, we do not observe large changes in $T_c$, Fig. 2 and supplementary information \cite{SI}.

The sample used in our experiments is a 5.1 mm long, 4 mm  diameter cylinder  of 98$\%$ porous silica aerogel. Following growth and supercritical drying, anisotropy was induced by axial compression of the sample by $19.4\%$. It had been used previously to study the field-temperature phase diagram of \she\ in compressed aerogel with magnetic impurities at high pressure (26 bar) \cite{Li.14b}, as well as to study the orientation of the orbital angular momentum relative to the anisotropy axis in the  \Bph\ \cite{Zim.18b}. Prior to compression, the same sample in its isotropic state was used to measure the phase diagram \cite{Pol.11, Li.13}. These experiments, carried out with magnetic impurity, provide an important baseline for the comparison with non-magnetic impurities that we report here.

In the present work, we performed measurements using pulsed NMR in magnetic fields ranging from $H = 49.1$ to $196$ mT with the field parallel to the aerogel anisotropy axis. The superfluid phases can be identified by the frequency shift, $\Delta\omega$, of the NMR resonance away from the Larmour frequency, $\omega_L$, as well as the magnetic susceptibility, $\chi$, which is proportional to the integral of the NMR spectrum. Measurements between 7.5 and 15 bar with magnetic impurities were taken to supplement earlier work  at 26 bar \cite{Li.14b,Li.15}. Then, sufficient \hefour\ to replace the solid \he\ on the surface, $\sim$3.5 layers, was mixed with \he\ at room temperature and introduced supercritically to the sample cell at $T > 10$\,K.  We verified the complete absence of solid \he\ on the aerogel surface using NMR. Measurements were conducted between 2.5 and 27 bar, during which the sample was warmed above 10\,K several times. There was no evidence for damage to the aerogel as might be indicated by a change in the normal state line width, nominally 5 ppm, or any change in the superfluid phase diagram.


\begin{figure}[t!]
\centerline{\includegraphics[width=0.45\textwidth]{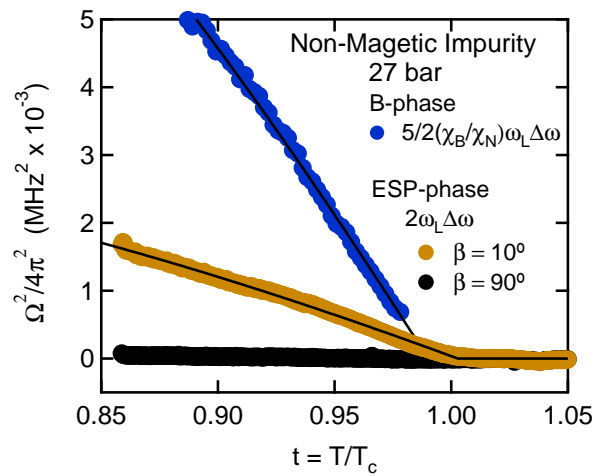}}
\caption{\label{fig:freq}(Color online) Longitudinal resonance frequencies taken at $P=27$\,bar with non-magnetic impurities, plotted versus reduced temperature, $t=T/T_c$. Data in the \Bph\ was taken at 0.1 T (blue circles) and data in the ESP phase at 0.2 T (yellow circles). Black circles are the frequency shift measured after a $90^\circ$ tip angle pulse in the ESP phase. The lack of frequency shift at $90^\circ$ is consistent with a Polar phase or 2D-disordered \Aph. Solid lines are fits used to extract the initial slope, as described in the text. }
\end{figure}


	The most striking result of  previous experiments on compressed silica aerogel with magnetic impurities is that the isotropic \Bph\ appears to be more stable than the anisotropic \Aph\ \cite{Li.14b, Li.15} in a small magnetic field.  This is contrary to theoretical predictions \cite{Thu.98, Vic.05, Aoy.06} which show that anisotropic scattering should stabilize anisotropic states. The phenomenon is manifest as a critical field, $H_c$, in the temperature-field phase diagram, Fig.\ref{fig:H_sq}a, which is defined by a tricritical point where \Aph\ and \Bph  coexist with the normal state. The tricritical point occurs at the intersection of the quadratic field dependent transition between the ESP and non-ESP phases ($A$ and $B$), $T_{AB}(H^2)$  with $T_c$. For an isotropic aerogel this intersection is precisely at $H=0$~\cite{Pol.11,Li.13}.  Moreover, we showed that $H_c$ is proportional to anisotropy based on a number of different compressed aerogel samples at $P=26$\,bar~\cite{Li.15}.  In each case the impurities were magnetic. The physical origin of the difference in free energy that favors the \Bph\ over the \Aph\ was not known.
	
	In the present work, at high pressure, we find that removing the magnetic impurity eliminates $H_c$ thereby demonstrating its origin, Fig.\ref{fig:H_sq}b. More precisely, it is the anisotropic distribution of magnetic impurities that gives rise to $H_c$.  Additionally, we extended measurements with magnetic impurities to lower pressure finding that $H_c$ decreases with decreasing pressure. It is essentially unmodified from 26 bar to 15 bar, reduced at 10 bar, and completely absent at 7.5 bar, Fig.\ref{fig:H_sq}a.  For the case of non-magnetic impurities, at low pressure an anisotropic ESP phase appears in a small window of temperature below $T_c$ in agreement with theoretical predictions which do not include magnetic quasiparticle scattering~\cite{Thu.98, Vic.05, Aoy.06}. The transition between the ESP and non-ESP phase is distinguished by the change in the temperature dependence of $\chi$. In the following we will describe how we have identified the ESP and non-ESP phases respectively as the {\it A} and {\it B} phases.

To identify the ESP and non-ESP phases, we look at the frequency shift, $\Delta\omega$, of the NMR resonance in the superfluid state which is dependent on the specific superfluid state, the orientation of the order parameter, and the tipping angle, $\beta$, of the NMR pulse~\cite{Hal.18}. The frequency shift determines the longitudinal resonance frequency, $\Omega$, which is proportional to the amplitude of the order parameter $\Delta$.

\begin{figure}
\centerline{\includegraphics[width=0.45\textwidth]{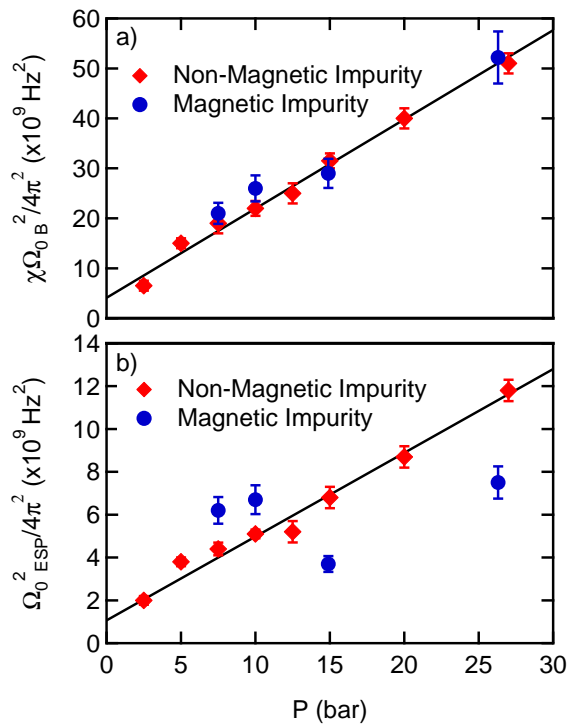}}
\caption{\label{fig:OmegaAB}(Color online). Initial slope of the temperature dependent longitudinal resonance frequency, $d\Omega^2/dt$  as a function of pressure: (a) in the \Bph, and (b) in the ESP phase. Blue circles (red diamonds) are measurements with (without) magnetic impurities. With non-magnetic impurity, both phases have a linear pressure dependence, an important indication of a common superfluid state throughout the range of pressure. The longitudinal resonance frequency is unchanged in the \Bph, but clearly changed in the ESP phase. Error bars are from fits as shown in Fig.\ref{fig:freq}.     }
\end{figure} 

We measured $\Delta\omega$ for the non-ESP phase with  magnetic and non-magnetic surface conditions and find that it has the same unique tip angle dependence, as the \Bph~\cite{Zim.18b}.  On this basis we identify the non-ESP phase as the \Bph.  At temperatures within 20\% of $T_c$ the angular momentum axis is perpendicular to the magnetic field resulting in a large frequency shift at small $\beta$ from which the longitudinal resonance frequency can be determined by,
\begin{align}
    \Omega_B^2(P,T)=\frac{5}{2}\omega_L\Delta\omega &&\beta\approx0^\circ.
\end{align}
\noindent The results are shown in Fig.\ref{fig:OmegaAB}a where we have multiplied $\Omega_B^2(P,T)$ by the magnetic susceptibility for later comparison with the ESP phase.

The \Bph\ longitudinal resonance frequency is temperature dependent, so we characterise it by the initial slope of $\left(\chi_B/\chi_N\right)\Omega_B^2$ relative to $T/T_c$ as $T$ approaches $T_c$, which we extract from a fit to the $T/T_c$ dependence of the frequency shift measured in pure superfluid \he\ \cite{Sch.93,Sch.92}. These fits are shown along with the data in Fig.\ref{fig:freq}. We denote this slope as $\chi\Omega^2_{0 B}=d((\chi_B/\chi_N)\Omega_B^2)/dt$. This quantity is plotted with and without magnetic impurities as a function of pressure in Fig.\ref{fig:OmegaAB}a. There is no discernible difference in the \Bph\ longitudinal resonance frequency with magnetic or non-magnetic scattering. We infer that, the \Bph\ order parameter is unaffected by the presence of magnetic impurities. Note that $\chi\Omega^2_{0 B}$ is linear in pressure. This linear pressure dependence is observed in pure \he\ \cite{Sch.93,Ran.96}, isotropic aerogel \cite{Pol.11}, and anisotropic aerogel \cite{Pol.12a}. It is a ubiquitous  property of \she\ phases~\cite{Zim.18}, and it is a useful measure of the uniformity of the superfluid state as a function of pressure.  We conclude that the non-ESP phase is the  \Bph\ at all pressures and is immune from magnetic impurity.

\begin{figure}
\centerline{\includegraphics[width=0.43\textwidth]{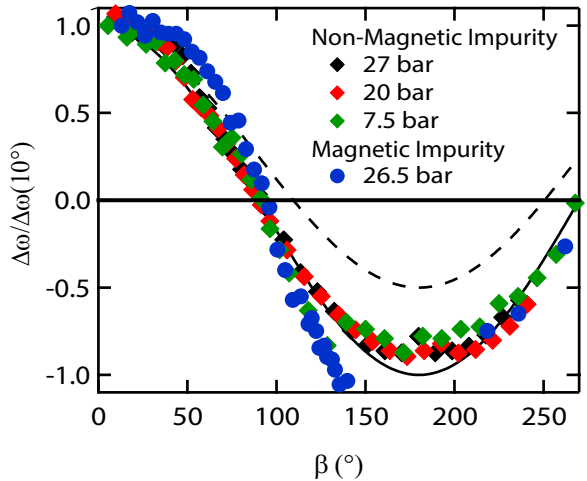}}%
\caption{\label{fig:cos}(Color online) Tip angle dependence of $\Delta\omega$ in the ESP phase at several pressures, with both magnetic and non-magnetic impurities. The solid curve is the calculated dependence for a Polar phase or 2D disordered \Aph, while the dashed line is for an ordered \Aph. In all cases the data agrees with the solid curve. }
\end{figure}

The identification of the ESP phase is more complicated. At high pressure in the same sample, we identified the ESP phase in the presence of magnetic impurity as the \Aph\ disordered into a  two-dimensional (2D) orbital glass, with its orbital angular momentum randomly oriented in the plane perpendicular to the aerogel anisotropy axis~\cite{Li.13}. This 2D glass phase was also seen in alumina aerogel~\cite{Ask.12,Dmi.15,Ask.15}, and its presence suggests that the nature of  disorder in axially compressed silica aerogel is the same as that 
of nematic aerogel.  The other candidate for the ESP phase is the Polar state. With magnetic field parallel to the aerogel anisotropy axis, both of these phases have identical tip angle dependence, with frequency shift given by,
\begin{equation}
2\omega_L\Delta\omega=\Omega^2_{ESP}(P,T)\cos(\beta),
\end{equation}
where $\Omega^2_{ESP}$ depends on the  superfluid state of the ESP phase and is larger for the Polar state than the \Aph\cite{Dmi.15,Zim.18}. At all pressures and impurities, the frequency shift in the ESP phase follows this behavior, as seen in Fig.\ref{fig:cos} and the tip angle dependence alone does not allow discrimination  between the two possible states.

More details can be obtained by investigating the magnitude of the longitudinal resonance frequency. Following the same procedure used for the $B$ phase, we can extract the initial slope of $\Omega^2_{ESP}$, which we denote as $\Omega^2_{0 ESP}=d(\Omega^2_{ESP})/dt$. This quantity is plotted as a function of pressure in Fig.\ref{fig:OmegaAB}. With non-magnetic impurities, $\Omega^2_{0 ESP}$ is linear in pressure, indicating that there is a single well-defined superfluid state throughout the whole pressure range. In contrast, with magnetic impurities, $\Omega^2_{0 ESP}$ behaves differently at high and low pressure. At high pressure, $\Omega^2_{0 ESP}$ is reduced by a factor of $\sim1.5$ by magnetic impurities, implying that the ESP phase is suppressed, while at low pressure $\Omega^2_{0 ESP}$ is larger than the value measured with non-magnetic impurities. We note that the transition between these two regions occurs between 10 and 15 bar, the same region where the critical field begins to decrease, Fig.\ref{fig:H_sq}. This change in behavior indicates that the ESP phase with magnetic impurities is a modified, or a different, superfluid state at low pressures. 
In either case, the results show that the ESP phase is strongly affected by magnetic impurity.

\begin{figure}[t]
\centerline{\includegraphics[width=0.43\textwidth]{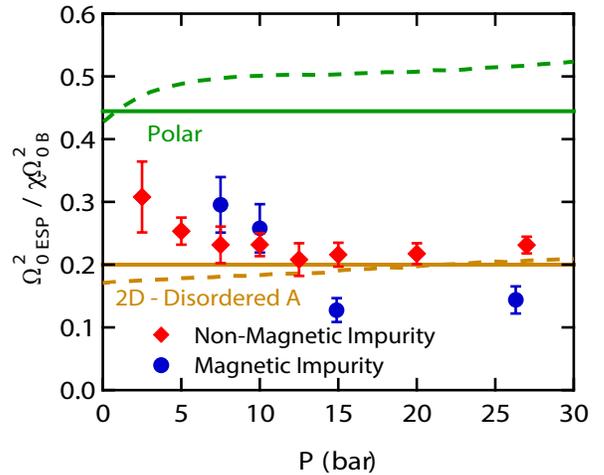}}
\caption{\label{fig:ABcomp}(Color online). $\Omega^2_{0 ESP}/\chi\Omega^2_{0 B}$ as a function of pressure with magnetic impurities (blue circles) and with non-magnetic impurities (red diamonds). Solid lines are the theoretical ratios discussed in the text between the 2D-disordered \Aph\ and \Bph\ (yellow), and the Polar phase and \Bph\ (green), while dashed lines are calculated from the pure \he\  Ginzburg Landau theory. At all pressures the data is more consistent with the 2D disordered \Aph, although the increase at low pressures may be due to Polar distortion. Error bars are calculated from the errors in $\Omega^2_{0 ESP}$ and $\chi\Omega^2_{0 B}$. }
\end{figure} 

A clear identification of the ESP phase requires a comparison of $\Omega^2_{0 ESP}$ with a known value. In nematic aerogel, the longitudinal resonance frequency of the \Aph\ of pure superfluid \he\ was used \cite{Dmi.18,Dmi.15}, due to the small amount of $T_c$ reduction in these samples. However, in silica aerogel, $T_c$ is suppressed by a larger amount, and the pure \he\ longitudinal resonance frequency has been found to be a poor reference point \cite{Thu.98,Pol.11,Pol.12a}.  Instead, we use $\Omega^2_B$ measured in the same aerogel sample for which the superfluid state is known. The ratio of longitudinal resonance frequencies of different phases is determined by the symmetry of those phases. The comparison with the \Bph\ has previously been used to identify the ESP phase as the \Aph\ in pure \he\ \cite{Osh.74b, Ran.94}, as well as in isotropic aerogel \cite{Pol.11}. We have calculated the ratio from our experimental values as $\Omega^2_{0ESP}/\chi\Omega^2_{0B}$, as shown in Fig.\ref{fig:ABcomp}.

For the 2D disordered \Aph, the ratio with the \Bph\ longitudinal resonance frequency is given by,
\begin{equation}
    \frac{\chi_N\Omega^2_A}{\chi_B\Omega^2_B}=\frac{1}{5}\left(\frac{\Delta_A}{\Delta_B}\right)^2,
\end{equation}
where $\Delta$ is the average amplitude of the order parameter~\cite{Zim.18, Leg.75}. We can take  $\Delta_A/\Delta_B\,\approx\,1$~\cite{Pol.11, Osh.74b}. Similarly, for the Polar phase, we have
\begin{equation}
    \frac{\chi_N\Omega^2_P}{\chi_B\Omega^2_B}=\frac{4}{5}\left(\frac{\Delta_P}{\Delta_B}\right)^2,
\end{equation}
where  we can use the  low pressure, weak coupling value of ($\Delta_P/\Delta_B)^2 = 5/9$~\cite{Zim.18}. These calculated ratios are shown as solid lines in Fig.\ref{fig:ABcomp}. Alternatively, $\Delta$ can be calculated from the experimental pure \he\  Ginzburg Landau parameters \cite{Cho.07} shown by the dashed lines in Fig.\ref{fig:ABcomp}.

Without  magnetic impurities, at high pressure, the experimental values of $\Omega^2_{0 ESP}/\chi\Omega^2_{0B}$ are consistent with the 2D disordered \Aph\ and rule out the Polar state. With magnetic impurities neither ratio is correct, indicating that the suppression of the \Aph\ distorts its order parameter changing the relative symmetry compared to the \Bph. At low pressure, both with and without magnetic impurities, $\Omega^2_{0ESP}/\chi\Omega^2_{0B}$  is larger than expected for the \Aph, though not as large as  for the Polar phase. This may be due to Polar distortion of the \Aph\ at low pressures, or a change in the \Aph\ itself. Further investigations are required to discriminate between these possibilities. 

In summary, the axially anisotropic, non-magnetic silica aerogel has two stable phases throughout the  pressure-temperature-field phase diagram which we have investigated. These are the $A$ and $B$ phases which preserve the same relative symmetry of the axial and isotropic states characteristic of pure \she.  This is also the case for magnetic impurity in isotropic aerogel \cite{Pol.11}. We have found that an anisotropic distribution of magnetic impurity is responsible for the critical field reported earlier~\cite{Li.14b,Li.15}, owing to the suppression of the $A$ phase order parameter.  This suppression leads to a violation of the relative symmetry relation between the two phases.  The \Bph\ is unaffected by magnetic impurities.

We are grateful to J. A. Sauls, J. J. Wiman, V. V. Dmitriev, and G. E. Volovik for helpful discussion, and support from the National Science Foundation (Grant No. DMR-1602542).

\bibliography{Zimref,PhysToday.Li}
\end{document}


\title{Supplementary Information: \\ ``The Effect of Magnetic Impurities on Superfluid \he\ in Aerogel''}

\author{A.M. Zimmerman}
\email[]{andrewzimmerman2016@u.northwestern.edu}
\author{M.D. Nguyen}
\author{J.W. Scott}
\author{W.P. Halperin}
\email[]{w-halperin@northwestern.edu}
\affiliation{Northwestern University, Evanston, IL 60208, USA}

\date{\today}

\maketitle

\begin{figure}

\centerline{\includegraphics[width=0.45\textwidth]{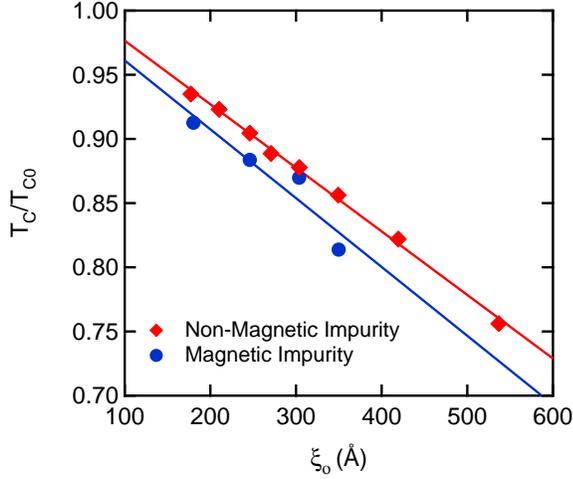}}

		\caption{\label{fig:Tc}(Color online). The ratio of $T_c$ in compressed aerogel to the pure superfluid value, $T_{c0}$, plotted versus the superfluid coherence length. This ratio is linearly dependent on the coherence length, both with and without magnetic impurities.}

\end{figure}

The energy scale of the superfluid transition temperature, $T_c$, is much larger than the energy difference between superfluid phases. Because of this, it requires significant effects from magnetic impurities to modify $T_c$. Dmitriev \et\ report a significant effect on $T_c$ in nematically ordered aerogel \cite{Dmi.18}. As noted in the main text, we find a much smaller change in $T_c$ in our less anisotropic silica aerogel, Fig. \ref{fig:Tc}. 

It is useful to consider the suppression of $T_c$ relative to its value in pure superfluid \he\, $T_{c0}$ , as, $T_c/T_{c0}$. The change in $T_c$ with the addition of magnetic impurities varies from $\approx2\%$ of $T_{c0}$ at high pressure to $\approx5\%$ of $T_{c0}$ at low pressure. Both with magnetic and with non-magnetic impurities, $T_c/T_{c0}$ is linear in the superfluid coherence length, $\xi_0$. We have previously discussed this relationship in our aerogel samples \cite{Zim.18b}. This linear dependence is characteristic of the effect of disorder on superfluidity and superconductivity, where $T_c$ suppression is determined by the impurity mean free path \cite{Abr.62,Lar.65,Sau.03,Fom.08}.  That this relationship holds both with and without magnetic scattering suggests that this description is also effective in the presence of magnetic impurities.

\bibliography{Zimref,PhysToday.Li}